\documentclass{article}
\usepackage{graphicx,amssymb,amsfonts,bm,mathptmx,enumerate}
\usepackage[tbtags]{amsmath}

\renewenvironment{figure}
{\par\medskip\refstepcounter{figure}\footnotesize\sffamily\selectfont}%
 {\par\medskip}

\renewcommand{\caption}[1] {\par{\footnotesize\sffamily Fig. 
\thefigure{}. #1}}

\begin{document}
\title{Casimir effect in external magnetic field}
\author{Marcin Ostrowski}
\date{}
\maketitle

\noindent
{\it Department of Theoretical Physics, University of Lodz,}\\
{\it ul.~Pomorska 149/153, 90-236 {\L}{\'o}d{\'z}, Poland}\\
{\it m.ostrowski@merlin.fic.uni.lodz.pl}

\begin{abstract}
In this paper we examine the Casimir effect for charged fields in presence of external magnetic field. We consider
scalar field (connected with spinless particles) and the Dirac field (connected with $1/2$-spin particles). In both cases we
describe quantum field using the canonical formalism. We obtain vacuum energy by direct solving field equations and using the
mode summation method. In order to compute the renormalized vacuum energy we use the Abel-Plana formula.
\end{abstract}

\section{Introduction}
The imposition of boundary condition on a quantum field leads to the modification of the vacuum energy level and can be
observed as an associated vacuum pressure. This effect (called Casimir effect) has been predicted by Casimir in his original
work in 1948\cite{int1}, and has been experimentally observed for electromagnetic field several years later. 
Until now, in many theoretical works Casimir energy has been computed for various types of boundary geometry and for fields
other than electromagnetic one. 

When we consider of charged fields another important question arise. How an external field, coupled to the charge,
affect the vacuum energy of the field? Answer to this question is important for understanding of some aspects in particle
physics. Within a hadron, for example, the vacuum energy of quark fields is affected by the electromagnetic field
of the quarks and by the color field of gluons and quarks.

In order to investigate how charged fermionic and bosonic constrained vacuum fluctuations are affected
by fields coupled to this charge, we consider vacuum energy of electrically charged fields under the influence of an
external constant uniform magnetic field, and constrained by simple boundary conditions.
In Section 2. we shall consider a complex scalar field confined between two infinite plates with Dirichlet boundary
conditions with magnetic field in a direction perpendicular to the plates. In \mbox{Section 3.} we shall consider a Dirac
quantum field under antiperiodic boundary conditions. This choice of geometry and external fields avoids technical
difficulties and focuses our attention on the fundamental issue.

The fermionic Casimir effect was first calculated by Johnson \cite{r6n4} for applications in the MIT bag model.
For a massless Dirac field, Johnson obtained an energy density 7/4 times the energy density of the
electromagnetic field.

In both cases (bosonic and fermionic) we describe quantum field using the canonical formalism. We obtain the vacuum energy
by direct solving the field equations and using the mode summation method. In order to compute the renormalized vacuum energy
we use the Abel-Plana formula.

In order to obtain influence of magnetic field on the Casimir energy another methods can be used. Recently, the papers
using Schwinger's proper time method have been published\cite{r6n1,r6n2}. So, our purpose is to compare results obtained
using our method with the Schwinger's one.

\section{Bosonic Casimir effect}

In this section we consider scalar field $\phi(\vec{r},t)\,$, describing charged, spinless particles with mass $m$.
It is under the influence of uniform magnetic field $\vec{B}=(0,0,B)$. We choose direction of $z$ axis in such a way
that $B$ is positive.

Using  gauge invariance, we choose electromagnetic potential corresponding to field $\vec{B}$ in the
form: $A_\mu=(0,-yB,0,0)$.

The lagrangian density of field $\phi$ takes the form:
\begin{multline}
\label{eq:Blag1}
\mathcal{L}=(D_\mu\phi)^*(D^\mu\phi)-m^2\phi^*\phi=\\
=\partial_t\phi^*\partial_t\phi-(\partial_x+ieBy)\phi^*(\partial_x-ieBy)\phi+\\
-\partial_y\phi^*\partial_y\phi-\partial_z\phi^*\partial_z\phi-m^2\phi^*\phi,
\end{multline}
and leads to the following equation of motion
\begin{equation}
\label{eq:Broro1}
\partial_t^2\phi-\partial_x^2\phi-\partial_y^2\phi-\partial_z^2\phi+2ieBy\partial_x\phi+\bigl(e^2B^2y^2+m^2\bigr){\phi}=0.
\end{equation}
where derivative $D_\mu=\partial_\mu+ieA_\mu$ in Eq.~\eqref{eq:Blag1} is covariant derivative.

Solution of Eq.~\eqref{eq:Broro1} is well known and is given in \cite{r6n3} in case of nonrelativistic quantum particle.
We briefly remind method of solving this kind of equation. The variables $x$, $z$ and $t$ do not occur in Eq.~\eqref{eq:Broro1}
explicity, therefore the solution of Eq.~\eqref{eq:Broro1} takes the form:
\begin{equation}
\label{eq:Bprzew1}
\phi(x,y,z,t)=F(y)e^{i(kx+pz-\omega{t})}.
\end{equation}
After inserting of Eq.~\eqref{eq:Bprzew1} into Eq.~\eqref{eq:Broro1} the equation for function $F(y)$ is given by:
\begin{equation}
\partial_y^2F+(\omega^2-k^2-p^2-m^2+2eBky-e^2B^2y^2)F=0,
\end{equation}
which has a solution in the form:
\begin{equation}
F(y)=e^{-eB/2(y-y_0)^2}\text{H}_n(\sqrt{eB}(y-y_0)),
\end{equation}
where H$_n$ are Hermite polynomials and $y_0=k(eB)^{-1}$.
The parameters $n$, $p$ i $\omega$ fulfil following relation
\begin{equation}
\omega=\sqrt{2eB(n+\tfrac{1}{2})+p^2+m^2}.
\end{equation}

Let us introduce the Dirichlet boundary conditions in the form:
\begin{equation}
\phi(x,y,z=0,t)=\phi(x,y,z=a,t)=0,
\end{equation}
which implies that field $\phi$ is equal to zero on the two parallel planes (plates). The plates are perpendicular to the
$z$-axis and the distance between them is equal to $a$.

In the area between plates, solution of Eq.~\eqref{eq:Broro1} is given by:
\begin{equation}
\phi(\vec{r},t)=\int_{-\infty}^\infty{dk}\sum_{l=1}^{\infty}\sum_{n=0}^\infty\bigl(a_{nlk}u_{nlk}(\vec{r})
e^{-i\omega_{nl}t}+b^+_{nlk}u_{nlk}(\vec{r})e^{i\omega_{nl}t}\bigr).
\end{equation}

Functions $u_{nlk}$ are special solutions of the Eq.~\eqref{eq:Broro1} in the form:
\begin{equation}
\label{eq:rospec1}
u_{nlk}(\vec{r})=C_ne^{ikx}\sin(p_lz)e^{-eB/2(y-y_0)^2}\text{H}_n(\sqrt{eB}(y-y_0)),
\end{equation}
where $C_n=(eBa^2)^{1/4}(2^nn!\pi^{3/2})^{-1/2}$ is normalisation constant, and $p_l=\pi{l}/a$.
Solutions of Eq.~\eqref{eq:rospec1} fulfil relations of orthogonality in the form:
\begin{equation}
\int_{\Gamma}d^3\vec{r}\,u_{n_1l_1k_1}(\vec{r})u_{n_2l_2k_2}^*(\vec{r})=a^2\delta(k_1-k_2)\delta_{l_1l_2}\delta_{n_1n_2}
\end{equation}
and relations of completeness in the form:
\begin{equation}
\int_{-\infty}^\infty{dk}\sum_{l=1}^\infty\sum_{n=0}^\infty{u}_{nlk}(\vec{r}_1)u_{nlk}^*(\vec{r}_2)=
a^2\delta^3(\vec{r}_1-\vec{r}_2),
\end{equation}
where $\Gamma$ is area between the plates.

Standard commutation relations for the field
\begin{eqnarray}
\,[\phi(\vec{r}_1,t),\partial_t\phi^+(\vec{r}_2,t)]&=&i\delta^3(\vec{r}_1-\vec{r}_2)\\
\,[\phi(\vec{r}_1,t),\phi(\vec{r}_2,t)]&=&[\partial_t\phi^+(\vec{r}_1,t),\partial_t\phi^+(\vec{r}_2,t)]=0
\end{eqnarray}
lead to the following commutators for creation and annihilation operators
\begin{multline}
\,[a_{n_1l_1}(k_1),a^+_{n_2l_2}(k_2)]=\,[b_{n_1l_1}(k_1),b^+_{n_2l_2}(k_2)]=
\tfrac{1}{2}a^{-2}\omega^{-1}_{n_1l_1}\delta_{n_1n_2}\delta_{l_1l_2}\delta(k_1-k_2).
\end{multline}

The hamiltonian density:
\begin{equation}
\mathcal{H}=\partial_t\phi^*\partial_t\phi+(\partial_x+ieBy)\phi^*(\partial_x-ieBy)\phi
+\partial_y\phi^*\partial_y\phi+\partial_z\phi^*\partial_z\phi+m^2\phi^*{\phi}
\end{equation}
leads to the vacuum energy in the form:
\begin{equation}
\label{eq:EvacB1}
E_{vac}=L^2\int_0^a{dz}\langle{0}{\vert}\mathcal{H}{\vert}0\rangle=L^2eB(2\pi)^{-1}
\sum_{l=1}^\infty\sum_{n=0}^\infty\sqrt{\tfrac{\pi^2}{a^2}l^2+eB(2n+1)+m^2},
\end{equation}
where $L$ is length of the plates.

The energy \eqref{eq:EvacB1} is infinite. Now, we define the renormalised vacuum energy $E_{ren}$ as a
difference between energy \eqref{eq:EvacB1} and energy without boundary conditions:
\begin{multline}
E_{ren}=L^2eB(2\pi)^{-1}\sum_{n=0}^\infty
\biggl(\sum_{l=1}^\infty\sqrt{\tfrac{\pi^2}{a^2}l^2+eB(2n+1)+m^2}+\\-
\int_{l=0}^\infty{dl}\sqrt{\tfrac{\pi^2}{a^2}l^2+eB(2n+1)+m^2}\biggr).
\end{multline}

In order to compute the renormalised energy we use the Abel-Plana formula in the form
\begin{equation}
\label{eq:ABF}
\sum_{l=1}^{\infty}f(l)-\int_{0}^{\infty}f(l)dl=-\tfrac{1}{2}f(0)
+i\int_{0}^{\infty}dt\frac{f(it)-f(-it)}{e^{2\pi{t}}-1}.
\end{equation}
This formula is often applied in renormalization problems. Its applications and generalizations can be found in many
works \cite{lit9,lit11}.

The renormalised energy, after using formula~\eqref{eq:ABF}, takes the final form:
\begin{equation}
\label{eq:ErenB1}
E_{ren}=-L^2{eB}a^{-1}\sum_{n=0}^\infty\int_{\lambda_n}^\infty{dt}\sqrt{t^2-\lambda_n^2}
\Bigl(e^{2\pi{t}}-1\Bigr)^{-1}
\end{equation}
where $\lambda_n=a\pi^{-1}\sqrt{eB(2n+1)+m^2}$.

Numerical calculations of Eq.~\eqref{eq:ErenB1} are shown in Figs.~1-5. Main conclusion is that, external magnetic field
decrease the Casimir energy and pressure of the scalar field (Fig.~1), and suppress it completely in the
limit $B\rightarrow0$. The decrease of the Casimir energy with a distance is faster for stronger magnetic field (Fig.~2).

Let us to compare the behaviour of the Casimir pressure for two different distances $a$ (Figs.~3,4). In order to suppress
the pressure noticeably, magnetic field should be of order of $10T$ for distance $a=10nm$, and of order of $0.1T$
for distance $a=100nm$.

Asymptotic behaviour of energy \eqref{eq:ErenB1} for strong magnetic field $(B\rightarrow\infty)$ we obtain
using the relation $\exp(2\pi{t})-1\rightarrow{\exp}(2\pi{t})$, which gives:
\begin{multline}
E_{B\rightarrow\infty}=
-L^2eBa^{-1}\sum_{n=0}^\infty\int_{\lambda_n}^\infty{dt}\sqrt{t^2-\lambda_n^2}\,e^{-2\pi{t}}=\\
=-\tfrac{1}{2}L^2eB\pi^{-2}\sum_{n=0}^\infty\sqrt{m^2+eB(2n+1)}\,\text{K}_1(2a\sqrt{m^2+2eB(2n+1)}),
\end{multline}
where $K_1(x)$ is Bessel function.

\begin{figure}
\begin{center}
\includegraphics[width=7cm]{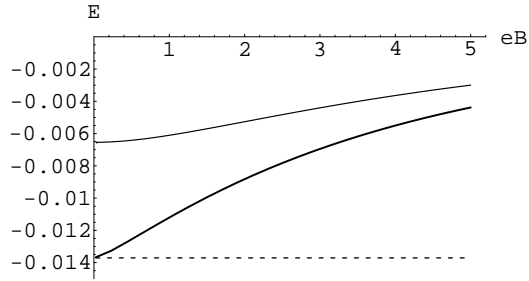}
\end{center}
\caption{Casimir energy for scalar field as a function of magnetic field $eB$, corresponding to $a=1$ and $L=1$. Bold curve
corresponds to mass $m=0$ whereas thin curve corresponds to mass $m=1$. The dashed line corresponds to value $-\pi^2/720$.
(It is standard value of energy for $B=0$)}
\end{figure}

\begin{figure}
\begin{center}
\includegraphics[width=7cm]{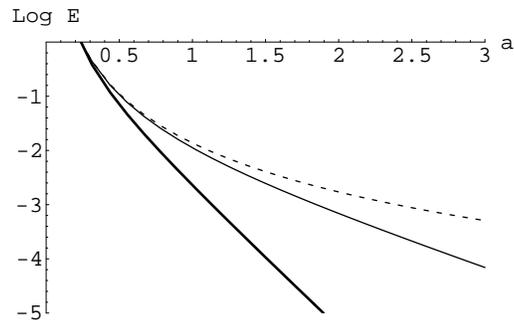}
\end{center}
\caption{Casimir energy for scalar field as a function of distance $a$, for massles field $m=0$ and $L=1$. Bold curve corresponds
to magnetic field $eB=8$ whereas thin corresponds to $eB=1$. The dashed curve corresponds to function $-\pi^2a^{-3}/720$
(standard function of distance $a$ for $B=0$). Values of energy are shown in logarithmic scale.}
\end{figure}

\begin{figure}
\begin{center}
\includegraphics[width=7cm]{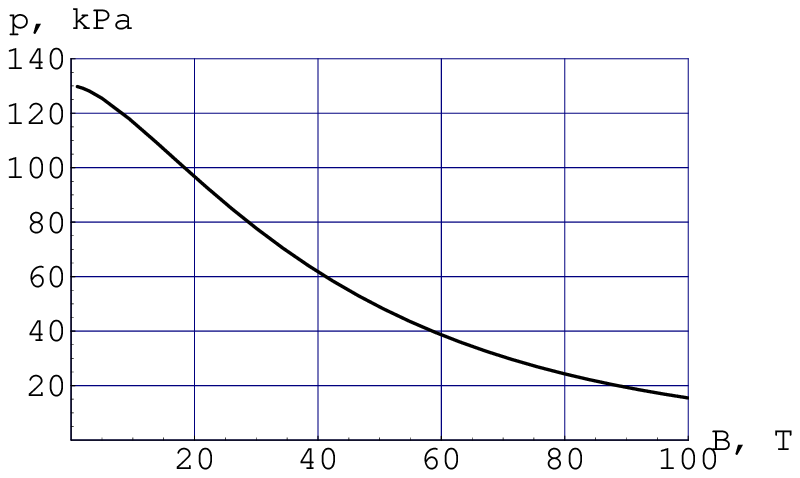}
\end{center}
\caption{Casimir pressure for scalar field as a function of magnetic field $B$, corresponding to distance $a=10nm$.
The diagram is made in SI units for massless field.}
\end{figure}

\begin{figure}
\begin{center}
\includegraphics[width=7cm]{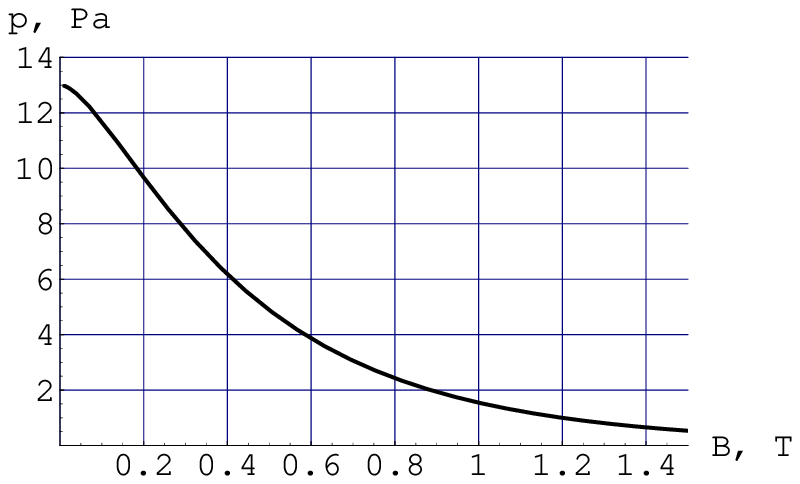}
\end{center}
\caption{Casimir pressure for scalar field as a function of magnetic field $B$, corresponding to distance $a=100nm$.
The diagram is made in SI units for massless field.}
\end{figure}

\begin{figure}
\begin{center}
\includegraphics[width=7cm]{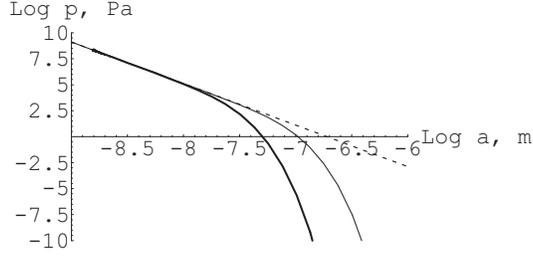}
\end{center}
\caption{Casimir pressure for massless, scalar field as a function of distance $a$. Diagram is made in SI units, using
logarithmic scales. Bold curve corresponds to $B=10T$, thin to $B=1T$, and dashed to $B=0T$.}
\end{figure}

Asymptotic behaviour of expression \eqref{eq:ErenB1} for weak magnetic field $(B\rightarrow0)$ we obtain, when we change
sum into integration, which gives:
\begin{multline}
\label{eq:ostatB1}
E_{B\rightarrow0}=-L^2eBa^{-1}\int_0^\infty{dn}\int_{\lambda(n)}^\infty{dt}\sqrt{t^2-\lambda^2(n)}
\Bigl(e^{2\pi{t}}-1\Bigr)^{-1}=\\
=-L^2a\pi^{-2}\int_1^\infty{dy}\sqrt{y^2-1}\int_M^\infty{dx}\,x^3(e^{2axy}-1)^{-1},
\end{multline}
where $M=\sqrt{m^2+eB}$. For $M=0$ Eq.~\eqref{eq:ostatB1} takes the standard value $E(0)=-L^2\pi^2/(720a^3)$
for charged, scalar field \cite{lit9}.

We compared Eq.~\eqref{eq:ErenB1} with Eq.~(15) from \cite{r6n1} (based on Schwinger's method) using {\it Mathematica 4.1}.
We obtained the same numerical results. However, we are not able analytically prove that these two equations are
equivalent.

\newpage
\section{Fermionic Casimir effect}

The lagrangian density of the Dirac field in external electromagnetic field takes the form:
\begin{equation}
\mathcal{L}=\Psi^+\gamma^0(i\gamma^\mu\partial_\mu-e\gamma^\mu{A_\mu}-m)\Psi,
\end{equation}
and leads to the following equation of motion
\begin{equation}
\label{eq:DirackB}
(i\gamma^\mu\partial_\mu-e\gamma^\mu{A}_\mu-m)\Psi^a=0,
\end{equation}
where $a=1,2,3,4$ are numbers of bispinor components.

We take $\gamma$ matrices in Dirac representation:
\begin{eqnarray}
\label{eq:gamma}
\gamma^0=
\begin{pmatrix}
I&0\\0&-I
\end{pmatrix},
\gamma^i=
\begin{pmatrix}
0&\sigma^i\\-\sigma^i&0
\end{pmatrix}.
\end{eqnarray}

Analogously as in previous section, we solve Eq.~\eqref{eq:DirackB} in presence of an external magnetic
field $\vec{B}=(0,0,B)$, but now we impose antiperiodic boundary conditions in the form:
\begin{equation}
\Psi(x,y,z=0,t)=-\Psi(x,y,z=a,t).
\end{equation}

Solution of Eq.~\eqref{eq:DirackB} can be predicted as follows:
\begin{equation}
\label{eq:predykcja}
\Psi^a(\vec{r},t)=\exp(i(\omega{t}+kx+pz))\eta^a(y),
\end{equation}
where bispinors $\eta^a(y)$ take the form:
\begin{equation}
\eta^a(y)=
\begin{pmatrix}
c_1u_{n-1}(y)\\
c_2u_n(y)\\
c_3u_{n-1}(y)\\
c_4u_n(y)\\
\end{pmatrix}
\end{equation}
where $c_a$ are number parameters independent of space co-ordinates, and functions $u_n(y)$ are given by:
\begin{equation}
\label{eq:fufu}
u_n(y)=e^{-eB/2(y-y_0)^2}\text{H}_n(\sqrt{eB}(y-y_0)),
\end{equation}
where $y_0=k/(eB)$.

After inserting Eq.~\eqref{eq:fufu} into Eq.~\eqref{eq:DirackB} we obtain matrix equation:
\begin{equation}
\label{eq:mac2B2}
\begin{pmatrix}
-(\omega+m)& 0& -p& 2n\sqrt{eB}\\
0& -(\omega+m)& \sqrt{eB}& p\\
p& -2n\sqrt{eB}& \omega-m& 0\\
-\sqrt{eB}& -p& 0& \omega-m 
\end{pmatrix}
\begin{pmatrix}
c_1\\c_2\\c_3\\c_4
\end{pmatrix}
=\vec{0}
\end{equation}
Nontrivial solutions of Eq.~\eqref{eq:mac2B2} (i.e. different from null vector) exist if rank of matrix is different
than 4. This situation occurs when:
\begin{equation}
\omega^2=p^2+2eBn+m^2.
\end{equation}

After inserting solutions of Eq.~\eqref{eq:mac2B2} into Eq.~\eqref{eq:predykcja} we obtain general solution of
Eq.~\eqref{eq:DirackB} in following form
\begin{multline}
\label{eq:rozw1B2}
\Psi(\vec{r},t)=\sum_{n=0}^\infty\sum_{l=-\infty}^\infty\int_{-\infty}^\infty{dk}\sum_s\Bigl(e^{-i(Et+kx+p_lz)})
a_{n,l,s}(k)\vec\xi_{n,l,s}(k,y)+
\\+e^{i(Et+kx+p_lz)})b^+_{n,l,s}(k)\vec\eta_{n,l,s}(k,y)\Bigr),
\end{multline}
where $p_l=\pi/a(2l+1)$, $E=\vert\omega\vert$, and variable $s$ takes the values -1, 1.

For $n>0$ bispinors $\vec\xi$, $\vec\eta$ from Eq.~\eqref{eq:rozw1B2} are given by:
\begin{eqnarray}
\vec\xi_{n,s=1}(k,p,y)&=&\frac{C_n}{\sqrt{2^{n-1}(n-1)!}}
\begin{pmatrix}
-\tfrac{p}{E+m}u_{n-1}(y)\\
\tfrac{\sqrt{eB}}{E+m}u_n(y)\\
u_{n-1}(y)\\
0
\end{pmatrix},\\
\vec\xi_{n,s=-1}(k,p,y)&=&\frac{C_n}{\sqrt{2^nn!}}  
\begin{pmatrix}
\tfrac{2n\sqrt{eB}}{E+m}u_{n-1}(y)\\
\tfrac{p}{E+m}u_n(y)\\
0\\
u_{n}(y)
\end{pmatrix},\\
\vec\eta_{n,s=1}(k,p,y)&=&\frac{C_n}{\sqrt{2^nn!}}
\begin{pmatrix}
0\\
u_n(y,-k)\\
-\tfrac{2n\sqrt{eB}}{m+E}u_{n-1}(y,-k)\\
\tfrac{p}{m+E}u_n(y,-k)
\end{pmatrix},\\
\vec\eta_{n,s=-1}(k,p,y)&=&\frac{C_n}{\sqrt{2^{n-1}(n-1)!}}
\begin{pmatrix}
u_{n-1}(y,-k)\\
0\\
-\tfrac{p}{m+E}u_{n-1}(y,-k)\\
-\tfrac{\sqrt{eB}}{m+E}u_n(y,-k)
\end{pmatrix},
\end{eqnarray}
where $C_n=\bigl((eB/\pi)^{1/2}(E+m)/2E\bigr)^{1/2}$.

For $n=0$ bispinors $\vec\xi_{0,s=-1}$ and $\vec\eta_{0,s=1}$ are also given by these equations, if we take
$u_{-1}(y)=0$. Whereas spinors $\vec\eta_{0,s=-1}$, $\vec\xi_{0,s=1}$ are equal to zero.

The spinors $\vec\eta$, $\vec\xi$ are orthonormal in the sense of the norm defined by:
\begin{equation}
\Vert\vec\xi_{n,s}(k,p,y)\Vert^2=\int_{-\infty}^\infty{dy}\vec\xi_{n,s}(k,p,y)\vec\xi^*_{n,s}(k,p,y).
\end{equation}

Anticommutation relations for the field operators $\Psi$, $\Pi=i\Psi^+$ in the form:
\begin{equation}
[\Psi_a(\vec{x},t),\Pi_b(\vec{y},t)]_+=i\delta(\vec{x}-\vec{y})\delta_{ab}
\end{equation}
lead to the following anticommutators for creation and annihilation operators
\begin{multline}
\,[a_{n_1l_1s_1}(k_1), a^+_{n_2l_2s_2}(k_2)]_+=[b_{n_1l_1s_1}(k_1), b^+_{n_2l_2s_2}(k_2)]_+=\\
=(2\sqrt{\pi}a)^{-1}\delta_{s_1s_2}\delta_{l_1l_2}\delta_{n_1n_2}\delta(k_1-k_2).
\end{multline}

In Dirac theory, the vacuum state is defined as the state with filled up negative energy levels. This situation
corresponds to:
\begin{align}
a_{nls}\vert{0}\rangle=&0 & b_{nls}\vert{0}\rangle=&0.
\end{align}

The hamiltonian density of the field:
\begin{equation}
\label{eq:B1gege}
\mathcal{H}=-\Psi^+\gamma^0(i\gamma^i\partial_i-e\gamma^\mu{A_\mu}-m)\Psi=i\Psi^+\partial_t\Psi
\end{equation}
(where in second step we use the fact, that $\Psi$ fulfils the Dirac equation) leads to the vacuum energy density as follows
\begin{multline}
\label{eq:wynikB2}
\mathcal{E}_{vac}(a)=\langle{0}\vert\mathcal{H}\vert{0}\rangle=\\
=\tfrac{-1}{2\sqrt{\pi}a}\sum_{n=0}^\infty\sum_{l=-\infty}^\infty\sum_s\int_{-\infty}^\infty{dk}\,E
\bigl(\vec\eta^*_{nl1}(k,y)\vec\eta_{nl1}(k,y)+\vec\eta^*_{nl,-1}(k,y)\vec\eta_{nl,-1}(k,y)\bigr)=\\
=-eB(\sqrt{\pi}a)^{-1}\sum_{n=0}^\infty\sum_{l=-\infty}^\infty{\alpha_n}\sqrt{\tfrac{\pi^2}{a^2}(2l+1)^2+2eBn+m^2}
\end{multline}
where $\alpha_0=1/2$ and $\alpha_n=1$ for $n>0$.

The energy density \eqref{eq:wynikB2} is infinite. Renormalised vacuum energy is difference between energy density
\eqref{eq:wynikB2} and vacuum energy without boundary conditions, and is given by:
\begin{multline}
\label{eq:ErenB2}
\mathcal{E}_{ren}(a)=-2eB(\sqrt{\pi}a)^{-1}\sum_{n=0}^\infty{\alpha_n}\Bigl(\sum_{l=0}^\infty
\sqrt{4\pi^2a^{-2}(l+1/2)^2+2eBn+m^2}+\\
-\int_0^\infty{dl}\sqrt{4\pi^2a^{-2}l^2+2eBn+m^2}\Bigr).
\end{multline}
In order to calculate Eq.~\eqref{eq:ErenB2} we use Abel-Plana formula in the usefull form:
\begin{equation}
\sum_{t=0}^\infty{F}(t+1/2)-\int_0^\infty{dt}F(t)=-i\int_0^\infty\frac{dt}{e^{2\pi{t}}+1}(F(it)-F(-it)),
\end{equation}
and finally we obtain renormalised energy density in the form:
\begin{equation}
\label{eq:ostatB2}
\mathcal{E}_{ren}(a)=-8eBa^{-2}\sum_{n=0}^\infty{\alpha_n}\int_{\lambda_n}^\infty\frac{dt}{e^{2\pi{t}}+1}
\sqrt{t^2-\lambda_n^2},
\end{equation}
where $\lambda_n=a(2\pi)^{-1}\sqrt{2eBn+m^2}$.

Numerical calculations of Eq.~\eqref{eq:ostatB2} are shown on Figs.~6,7.
In contrast to the bosonic case, for the Dirac field the Casimir energy is enhanced by the external magnetic field (Fig.~6).
So, the Dirac vacuum behaves like paramagnetic medium. Decrease of the Casimir energy with a distance is slower in strong
magnetic field (Fig.~7).

\begin{figure}
\begin{center}
\includegraphics[width=7cm]{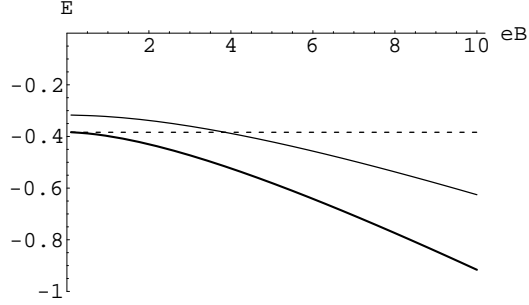}
\end{center}
\caption{Casimir energy for the Dirac field as a function of magnetic field $eB$, corresponding to distance $a=1$ and $L=1$.
Bold curve corresponds to mass $m=0$ whereas thin corresponds to mass $m=1$. Dashed line corresponds to value
equal to $-7\pi^2/180$ (standard value of energy for $B=0$).}
\end{figure}

\begin{figure}
\begin{center}
\includegraphics[width=7cm]{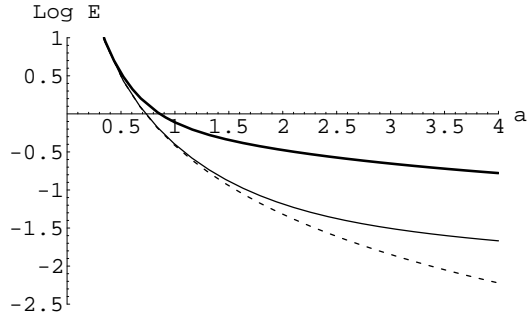}
\end{center}
\caption{Casimir energy for the Dirac field as a function of distance $a$, for massles field $m=0$ and $L=1$. Bold curve corresponds
to magnetic field $eB=8$ whereas thin corresponds to $eB=1$. Dashed curve corresponds to function $-7\pi^2a^{-3}/180$
(standard function of energy for $B=0$). Values of energy are shown in logarithmic scale.}
\end{figure}

Vacuum pressure is shown on Figs.~8-9. For massless field (and strong $B$) it is increasing linear function
of magnetic field $B$ (Fig.~8). For massive field, pressure depends on the field $B$ very weakly - two curves on Fig.~9
(for $B=1T$ and $B=100T$) coincide. For electron field values of Casimir pressure are measurable for distances
smaller then $0.1nm$ (Fig.~9). In practice it excludes possibilities of experimental tests. It can play a role only in
subatomic distances, for example in some particle models.

Analogously as for bosons, we compared Eq.~\eqref{eq:ostatB2} with Eq.~(9) from \cite{r6n2} (based on Schwinger's method)
using {\it Mathematica 4.1}. We obtained the same numerical results. However, we are not able analytically prove that these
two equations are equivalent.

\begin{figure}
\begin{center}
\includegraphics[width=7cm]{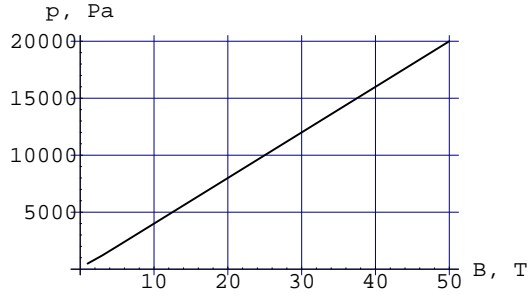}
\end{center}
\caption{Casimir pressure for massles, the Dirac field as a function of magnetic field $B$, corresponding to
distance $a=100nm$. The diagram is made in SI units.}
\end{figure}

\begin{figure}
\begin{center}
\includegraphics[width=7cm]{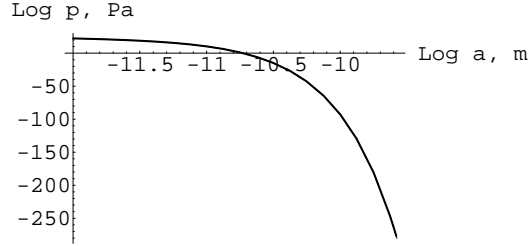}
\end{center}
\caption{Casimir pressure for the Dirac field of electrons ($m=m_e$) as a function of distance $a$. The diagram
is made in SI units for two values of field $B=1T$ and $B=100T$. Both curves coincide.}
\end{figure}

Asymptotic behaviour of energy density \eqref{eq:ostatB2} for strong magnetic field $(B\rightarrow\infty)$ and for
$a\,m\rightarrow\infty$ we obtain using the relation $\exp(2\pi{t})-1\rightarrow{\exp}(2\pi{t})$, what gives:
\begin{multline}
\mathcal{E}_{B\rightarrow\infty}=
-8{eB}a^{-2}\sum_{n=0}^\infty{\alpha_n}\int_{\lambda_n}^\infty{dt}\sqrt{t^2-\lambda_n^2}\,e^{-2\pi{t}}=\\
=-2eB\pi^{-2}a^{-1}\sum_{n=0}^\infty{\alpha_n}\sqrt{m^2+2eBn}\,\text{K}_1(a\sqrt{m^2+2eBn}).
\end{multline}

Asymptotic behaviour of Eq.~\eqref{eq:ostatB2} for weak magnetic field $(B\rightarrow0)$ we obtain changing sum into
integration, which gives:
\begin{multline}
\label{eq:zeroB2}
\mathcal{E}_{B\rightarrow0}=-8eBa^{-2}\int_0^\infty{dn}\int_{\lambda(n)}^\infty{dt}\sqrt{t^2-\lambda^2(n)}
\Bigl(e^{2\pi{t}}+1\Bigr)^{-1}=\\
=-2\pi^{-2}\int_1^\infty{dy}\sqrt{y^2-1}\int_m^\infty{dx}\,x^3(e^{2axy}+1)^{-1},
\end{multline}
For $m=0$ expression \eqref{eq:zeroB2} is equal to $\mathcal{E}(0)=-7\pi^2/(180a^4)$, which is standard value of Casimir energy
of the Dirac field without external fields \cite{r6n4}.
The expression \eqref{eq:zeroB2} donot depends on $B$, which suggests that $\mathcal{E}_{B\rightarrow0}=\mathcal{E}_{(B=0)}+\mathcal{O}(B^2)$.

\section*{Acknowledgements}
I wish to thank B.~Broda for numerous discussions about results.

\noindent This article was supported by the Polish Ministry of Scientific Research and Information Technology under
the grant No. PBZ/MIN/008/P03/2003 and by the University of Lodz.

\end{document}